\pdfoutput=1

\documentclass[11pt]{article}

\usepackage[final]{acl}

\usepackage{times}
\usepackage{latexsym}

\usepackage[T1]{fontenc}

\usepackage[utf8]{inputenc}

\usepackage{microtype}

\usepackage{inconsolata}

\usepackage{graphicx}

\usepackage{bm}
\usepackage{algorithm,algorithmic}
\usepackage{amsfonts,amsmath,amssymb,amsthm,mathrsfs}
\usepackage{listings}
\usepackage[dvipsnames]{xcolor}

\title{WaveFM: A High-Fidelity and Efficient Vocoder Based on Flow Matching}

\author{Tianze Luo, Xingchen Miao, Wenbo Duan \\
    Tsinghua Universtiy\\
    \texttt{\{ltz22,miuxc22,dwb22\}@mails.tsinghua.edu.cn}
  }

\begin{document}
\maketitle
\begin{abstract}
Flow matching offers a robust and stable approach to training diffusion models. However, directly applying flow matching to neural vocoders can result in subpar audio quality. In this work, we present WaveFM, a reparameterized flow matching model for mel-spectrogram conditioned speech synthesis, designed to enhance both sample quality and generation speed for diffusion vocoders. Since mel-spectrograms represent the energy distribution of waveforms, WaveFM adopts a mel-conditioned prior distribution instead of a standard Gaussian prior to minimize unnecessary transportation costs during synthesis. Moreover, while most diffusion vocoders rely on a single loss function, we argue that incorporating auxiliary losses, including a refined multi-resolution STFT loss, can further improve audio quality. To speed up inference without degrading sample quality significantly, we introduce a tailored consistency distillation method for WaveFM. Experiment results demonstrate that our model achieves superior performance in both quality and efficiency compared to previous diffusion vocoders, while enabling waveform generation in a single inference step.
\footnote{Our codes are available at \url{https://github.com/luotianze666/WaveFM}.}

\end{abstract}

\section{Introduction}
\label{sec:intro}

Recent advancements in network architectures and training algorithms have greatly enhanced the ability of deep generative models to produce high-fidelity audio in speech synthesis \citep{lee2021priorgrad,lee2022bigvgan,siuzdak2023vocos,huang2023fastdiff,
wang2023neural,nguyen2024fregrad,ju2024naturalspeech,kumar2024high}. The initial breakthrough came with the autoregressive generation of waveforms from mel-spectrograms \citep{oord2016wavenet,kalchbrenner2018efficient}, which provided high audio fidelity but suffered from slow inference speeds. To enable real-time high-fidelity speech synthesis, a variety of non-autoregressive models have been introduced, classified broadly into three categories: flow-based models, generative adversarial networks (GANs), and diffusion models.

Flow-based models utilize invertible neural networks to generate waveforms from a selected prior distribution, such as a Gaussian distribution, estimating log-likelihoods during training \citep{ping2020waveflow,prenger2019waveglow}. While these intricately designed models maintain invertibility and evaluate determinants, this complexity limits their flexibility, and consequently the quality of the audio output.

Generative Adversarial Networks (GANs) provide greater flexibility than flow-based models and can generate waveforms with high fidelity more efficiently \citep{kumar2019melgan,kong2020hifi,kim2021fre,jang2021univnet,lee2022bigvgan,siuzdak2023vocos}. Their success stems from the generators' large receptive fields and the discriminators' ability to detect noise across various scales and periods. For instance, \citet{kumar2019melgan} introduced multi-scale discriminators, while \citet{kong2020hifi} developed a multi-receptive field (MRF) generator alongside multiple multi-period discriminators, leading to substantial improvements. Moreover, \citet{lee2022bigvgan} further enhanced sample quality by utilizing the snake activation function and integrating the anti-aliased multi-periodicity (AMP) composition module.

Denoising diffusion probabilistic models (DDPMs) have recently gained significant popularity for their ability to transform a simple prior distribution into a complex ground truth distribution through a Markov chain process \citep{kong2020diffwave,lam2022bddm,huang2023fastdiff,nguyen2024fregrad}. These models rely on a parameter-free noise-adding diffusion process to generate training data for the denoising generator, eliminating the need for auxiliary networks like discriminators or autodecoders during training. However, the inference phase of diffusion models tends to be time-consuming. To mitigate this issue, \citet{kong2020diffwave}, \citet{lam2022bddm}, and \citet{huang2023fastdiff} introduced several fast-sampling algorithms that speed up waveform generation, though with a minor compromise in sample quality. Consistency models \citep{song2023consistency,song2023improved} have been proposed to enhance the efficiency of diffusion models by directly predicting the endpoint of the probability flow ordinary differential equation (PF-ODE) at each step, enabling single-step inference. These models surpass previous distillation approaches in image synthesis tasks and achieve higher distillation efficiency by aligning along ODE trajectories, thereby avoiding the need to numerically solve the entire ODE.

In this study, we propose WaveFM for the mel-spectrogram conditioned speech synthesis task. Firstly, since the mel-spectrogram records the energy information in waveforms, an appropriately conditioned distribution can significantly improve sample quality. Additionally, we adopt a reparameterized flow matching method that directly predicts the waveform, allowing us to apply several auxiliary losses to the original flow-matching loss to further enhance the model's performance. 

To better supervise the model's output waveforms regarding phase angles, we also incorporate a multi-resolution phase loss into our model.
Furthermore, gradient and Laplacian operators are utilized on the real and generated spectrograms. Minimizing the mean square losses associated with these operations enables the model to better learn edge details and structural patterns in the spectrograms.

Finally, we propose a tailored consistency distillation method for WaveFM to further accelerate the model's inference speed while maintaining audio quality.
The subjective and objective experiment results indicate that WaveFM outperforms previous diffusion models in terms of sample quality and efficiency, and generalizes better on out-of-distribution musical mel-spectrograms.

\section{Related Works}

\subsection{Flow Matching and Rectified Flow Models}
\label{rw}
Flow matching \citep{lipman2022flow} and rectified flow \citep{liu2022flow} models share a similar training objective, and diffusion models can also be interpreted within this framework. Here, we briefly introduce their mathematical principles using simpler, self-contained notation.

\paragraph{Theorem 1} Let $\bm{x}_t$ be a continuously differentiable random process on $t \in [0,1]$ and $p(\bm{x},t)$ be its probability density function. We denote the prior distribution as $\bm{x}_0$ and the ground truth distribution as $\bm{x}_1$. If the conditional expectation
$\mathbb{E}\left[\frac{\operatorname{d}\!\bm{x}_t}{\operatorname{d}\!t}\Big|\bm{x}_t=\bm{x}\right]$
is locally Lipschitz, we let
\begin{equation}
\bm{v}(\bm{x},t)=\mathbb{E}\left[\frac{\operatorname{d}\!\bm{x}_t}{\operatorname{d}\!t}\bigg\vert
\bm{x}_t=\bm{x}\right].
\end{equation}
Then samples from the data distribution $\bm{x}_1$ can be obtained by sampling from the prior distribution $\bm{x}_0$ and solving the following ODE with an initial value $\bm{x}_0$ at time $t=0$:
\begin{equation}
\frac{\operatorname{d}\!\bm{x}}{\operatorname{d}\!t}=\bm{v}(\bm{x},t).
\end{equation}

The detailed proof is available in \autoref{sec:appa}. The conditional expectation can be expressed as a simple mean square training objective:
\begin{equation}
\min\limits_{\bm{v}}\mathbb{E}\left\lVert
\frac{\operatorname{d}\!\bm{x}_t}{\operatorname{d}\!t}-\bm{v}(\bm{x}_t,t)\right\rVert^2.
\end{equation}

In practice, straight trajectories generally imply lower transportation costs. Thus, we take
\begin{equation}
\bm{x}_t=t\bm{x}_1+(1-t)\bm{x}_0, \quad t \in [0,1],
\end{equation}
which leads to the following objective for the neural network $\bm{v}(\bm{x},t)$. After the training process, data samples can be generated by numerically solving the ODE according to \textbf{Theorem 1}.
\begin{equation}
\label{origintarget}
\min\limits_{\bm{v}}\mathbb{E}\left\lVert\bm{x}_1-\bm{x}_0-\bm{v}(\bm{x}_t,t)\right\rVert^2.
\end{equation}

\subsection{Consistency Distillation}

Consistency Distillation (CD) \citep{song2023consistency} is an efficient method for distilling diffusion models to enable one-step generation. In the original paper, the authors adopt the following forward Stochastic Differential Equation (SDE) to diffuse data:
\begin{equation}
    \operatorname{d}\!\bm{x} = \sqrt{2t} \operatorname{d}\!\bm{w}, \quad t \in [\epsilon,T],
\end{equation}
where $\epsilon=0.002$ and $T=80$. The corresponding backward SDE and PF-ODE are given by:
\begin{equation}
    \operatorname{d}\!\bm{x} = - 2t\nabla_{\bm{x}}\log p(\bm{x},t) \operatorname{d}\!\bm{t} + \sqrt{2t} \operatorname{d}\!\bm{w},
\end{equation}
\begin{equation}
    \operatorname{d}\!\bm{x} = - t\nabla_{\bm{x}}\log p(\bm{x},t) \operatorname{d}\!\bm{t}. 
\end{equation}

In CD, the time steps are discretized by
\begin{equation}
    t_i = \epsilon^{1/\rho} + \frac{i-1}{N-1} (T^{1/\rho} - \epsilon^{1/\rho}), 
\end{equation}
where $N$ is the total number of discretization steps, $\rho = 7$, and $i \in \{1, \cdots, N\}$. According to the forward SDE, CD samples $n \sim \mathcal{U}\{1,2,\dots,N-1\}$, $\bm{x}_{t_{n+1}} \sim \mathcal{N}\left(\bm{x}_{\text{clean}}, t_{n+1}^2 \mathbf{I}\right)$, and then the pretrained teacher score network is used to compute $\hat{\bm{x}}_{t_{n}}$ by numerically solving the PF-ODE, where any type of ODE solver can be chosen for this purpose. The student network in CD aims to predict the endpoint at time $t=\epsilon$ of the PF-ODE trajectory at any position and time, parameterized as follows:
\begin{equation}
    \bm{f}_{\bm{\theta}}(\bm{x},t)=c_{\text{skip}}(t)\bm{x}+c_{\text{out}}(t)\bm{F}_{\bm{\theta}}(\bm{x},t),
\end{equation}
where $c_{\text{skip}}(t)=\frac{\sigma^2_{\text{data}}}{(t-\epsilon)^2+\sigma^2_{\text{data}}}$,
$c_{\text{out}}(t)=\frac{\sigma_{\text{data}}(t-\epsilon)}{\sqrt{\sigma^2_{\text{data}}+t^2}}$, $\sigma_{\text{data}}=0.5$
and $\bm{F}_{\bm{\theta}}(\bm{x},t)$ is the neural network.
The loss function of CD is given by
\begin{equation}
    \lambda(t_n) d\left(\bm{f}_{\bm{\theta}}(\bm{x}_{t_{n+1}},t_{n+1}),
    \bm{f}_{\bm{\theta}^{-}}(\hat{\bm{x}}_{t_{n}},t_{n})\right),
\end{equation}
where $\lambda(t_n)$ is a scale function, and $d(\cdot,\cdot)$ is a distance function such as L2 distance. The parameters $\bm{\theta}^{-}$ are updated using Exponential Moving Average (EMA), where $\mu$ is the EMA decay rate.
\begin{equation}
    \bm{\theta}^- \leftarrow \operatorname{stopgrad}(\mu \bm{\theta}^- + (1-\mu)\bm{\theta}).
\end{equation}

\section{Methodology}
\subsection{Mel-Conditioned Prior Distribution}

Mel-conditioned prior distributions have been applied to diffusion models \citep{lee2021priorgrad,koizumi2022specgrad}, but they do not closely approximate the audio distribution due to the necessity of stabilizing diffusion training objectives. For instance, \citet{lee2021priorgrad} utilize $\mathcal{N}(\bm{\mu},\bm{\Sigma})$ as the diffusion prior distribution, with their training objective defined as
\begin{equation}
   \bm{x}_t = \sqrt{\bar{\alpha}_t}(\bm{x}_0 - \bm{\mu}) + \sqrt{1 - \bar{\alpha}_t^2}\bm{\epsilon},
\end{equation}
\begin{equation}
   \min\limits_{\bm{\epsilon}} 
   (\bm{\epsilon} - \bm{\epsilon}_{\bm{\theta}}(\bm{x}_t,t))^\top
   \bm{\Sigma}^{-1}
   (\bm{\epsilon} - \bm{\epsilon}_{\bm{\theta}}(\bm{x}_t,t)),
\end{equation}
where $\bm{x}_0 \sim p_{data}$ and $\bm{\epsilon} \sim \mathcal{N}(\mathbf{0},\bm{\Sigma})$. They set $\bm{\mu} = \bm{0}$ and $\bm{\Sigma}$ as a diagonal matrix derived from the mel-spectrogram. Nonetheless, to stabilize the training process, they need to clamp the standard deviations between 0.1 and 1, which increases the distance between the prior and the audio distribution.

According to \textbf{Theorem 1}, however, to train a flow matching model, we only need to sample from two marginal distributions without requiring their analytical forms. This means that WaveFM could utilize a prior distribution with much smaller variance without compromising training stability. We choose $\mathcal{N}(\mathbf{0},\bm{\Sigma})$ as the prior distribution with a diagonal $\bm{\Sigma}$. We utilize the logarithmic mel-spectrograms as inputs to the neural network, as the raw values span a wide range of $[0, 32768]$. Since the mel-spectrogram captures the energy of the audio signal, the square root of the sum across the frequency dimension is a suitable choice for the standard deviation of the prior distribution. We normalize it by dividing it by $\sqrt{\texttt{mel-bands} \times 32768}$ to ensure that it falls within $[0,1]$, apply linear interpolation to align its shape with the audio, and clamp the values with a minimum of $10^{-3}$. Given that the values in a mel-spectrogram are typically much smaller than the potential maximum value, the standard deviation can indeed approach $10^{-3}$ in nearly silent regions. This suggests that our prior distribution is aligned with the audio distribution more closely. Our ablation study indicates that the adopted prior distribution enhances the sample quality of WaveFM.

\subsection{Training Objective}
We follow the notation in \autoref{rw}.
The original objective in \autoref{origintarget} aims to estimate a random derivative, which not only prevents the incorporation of auxiliary losses, such as the mel-spectrogram loss, but also complicates the design of a neural network with periodic inductive bias, as the random noise can disrupt periodic patterns. Experiment results demonstrate that this original objective can result in inferior sample quality for our network. Therefore, we hope the neural network to directly generate audio from noise, rather than predicting random derivatives. To achieve this, we choose to reparameterize the original mean square objective as follows:
\begin{equation}
    \bm{x}_1-\bm{x}_0=\frac{\bm{x}_1-\bm{x}_t}{1-t},\quad t\in [0,1).
\end{equation}
\begin{equation}
    \bm{v}(\bm{x}_t,t)=\mathbb{E}[\bm{x}_1-\bm{x}_0\vert\bm{x}_t]=\frac{\mathbb{E}[\bm{x}_1|\bm{x}_t]-\bm{x}_t}{1-t}.
\end{equation}
\begin{equation}
    \Leftrightarrow \min\limits_{\bm{v}^\prime} \frac{\mathbb{E}\left\lVert\bm{x}_1-\bm{v}^\prime (\bm{x}_t,t)\right\rVert^2}{1-t},t\in [0,1).
\end{equation}

We can now directly utilize a neural network to predict clean audio from mel-spectrograms, similar to GANs, allowing for the straightforward addition of auxiliary losses to the mean square loss:
\begin{equation}
\begin{aligned}
    \min\limits_{\bm{v}'}\Big(&\frac{1}{1-t}\mathbb{E}\left\lVert\bm{x}_1-\bm{v}^\prime (\bm{x}_t,t)\right\rVert^2\\
    +&\lambda_0\mathbb{E}\left[\operatorname{STFTLoss}(\bm{x}_1,\bm{v}^\prime(\bm{x}_t,t))\right]\\
    +&\lambda_1\mathbb{E}\frac{\left\lVert\operatorname{mel}(\bm{x}_1)-\operatorname{mel}(\bm{v}^\prime (\bm{x}_t,t))\right\rVert_1}{M}\Big).
\end{aligned}
\end{equation}

The total loss function employed by WaveFM is defined as above, with $\lambda_0=0.02$, $\lambda_1=0.02$, and $M$ is the number of elements in the
mel-spectrograms.
To prevent factor \(\frac{1}{1-t}\) from approaching infinity and stabilize training, we set the coefficient to 10 for \(t \in [0.9,1)\). 
The first auxiliary loss function is a multi-resolution STFT loss, initially introduced by Parallel WaveGAN \citep{yamamoto2019parallel}. They apply the short-time Fourier transform (STFT) at three resolutions to both the clean audio and the generated audio, with FFT, hop, and window sizes set to $(1024, 2048, 512)$, $(120, 240, 50)$, and $(600, 1200, 240)$, respectively. The spectral convergence loss $L_{\text{sc}}$ and log STFT magnitude loss $L_{\text{mag}}$ are computed as follows:
\begin{equation}
    L_{\text{sc}}(\bm{x},\hat{\bm{x}})=\frac{\left\lVert\left\lvert\operatorname{STFT}(\bm{x})\right\rvert-\left\lvert\operatorname{STFT}(\hat{\bm{x}})\right\rvert\right\rVert_F}{\left\lVert |\operatorname{STFT}(\bm{x})| \right\rVert_F},
\end{equation}
\begin{equation}
    L_{\text{mag}}(\bm{x},\hat{\bm{x}})=\frac{1}{N}\left\lVert\log\frac{\left\lvert\operatorname{STFT}(\bm{x})\right\rvert}{\left\lvert\operatorname{STFT}(\hat{\bm{x}})\right\rvert}\right\rVert_1,
\end{equation}
where $\bm{x},\hat{\bm{x}}$ denote the clean and generated audios, respectively; $N$ is the number of elements in the STFT spectrograms; $\lVert\cdot\rVert_F, \lVert\cdot\rVert_1$ denote the Frobenius and L1 norms; operators to the STFT-shaped matrices inside the norms are element-wise. Thus, the total loss function equals
\begin{equation}
    \operatorname{STFTLoss}(\bm{x},\hat{\bm{x}})= \frac{1}{3} \sum\limits_{m=1}^3 (L^{(m)}_{\text{sc}}+L^{(m)}_{\text{mag}})(\bm{x},\hat{\bm{x}})
\end{equation}

Notably, the original multi-resolution STFT loss leverages only the magnitude information in STFT spectrograms. So we replace $L_{\text{sc}}$ with a phase angle loss $L_{\text{pha}}$, defined as
\begin{equation}
    \Delta P = \operatorname{Phase}(\operatorname{STFT}(\bm{x}))-\operatorname{Phase}(\operatorname{STFT}(\hat{\bm{x}})),
\end{equation}
\begin{equation}
    L_{\text{pha}}(\bm{x},\hat{\bm{x}})=\frac{\left\lVert
    \operatorname{atan2}(\sin\Delta P,\cos\Delta P)\right\rVert_1}{N},
\end{equation}
where $\operatorname{atan2}$ is used to wrap the phase difference into $(-\pi,\pi]$. We do not compute the phase angle loss where the squared magnitude is less than $1 \times 10^{-6}$, as the phase angles there are insignificant and can produce excessively large gradients that destabilize the model. In our implementation, we add a small constant of $1 \times 10^{-6}$ to the squared magnitude for the computation of $L_{\text{mag}}^{(m)}$, and we adjust the FFT, hop, and window sizes to $(1024, 2048, 512)$, $(128, 256, 64)$, and $(512, 1024, 256)$, respectively.

Finally, to further enhance the detection capability of our multi-resolution STFT loss, we apply temporal gradient, frequency gradient, and Laplacian operators to the magnitude of both the clean and generated spectrograms. These operators are defined respectively as follows:  
\begin{equation}\footnotesize  
    \frac{1}{4}\begin{bmatrix}  
        -1&1\\-2&2\\-1&1  
    \end{bmatrix},  
    \frac{1}{4}\begin{bmatrix}  
        -1&-2&-1\\1&2&1  
    \end{bmatrix},  
    \frac{1}{8}\begin{bmatrix}  
        -1&-1&-1\\-1&8&-1\\-1&-1&-1  
    \end{bmatrix}.  
\end{equation}  
We then compute their mean square errors, scaled by 4, 4 and 2, respectively, to help the model learn the regular patterns within the spectrograms. These operators enhance the visibility of edge information in the spectrograms, enabling the model to capture finer details more effectively.  

\begin{figure}[h]
    \centering
    \includegraphics[width=1\linewidth]{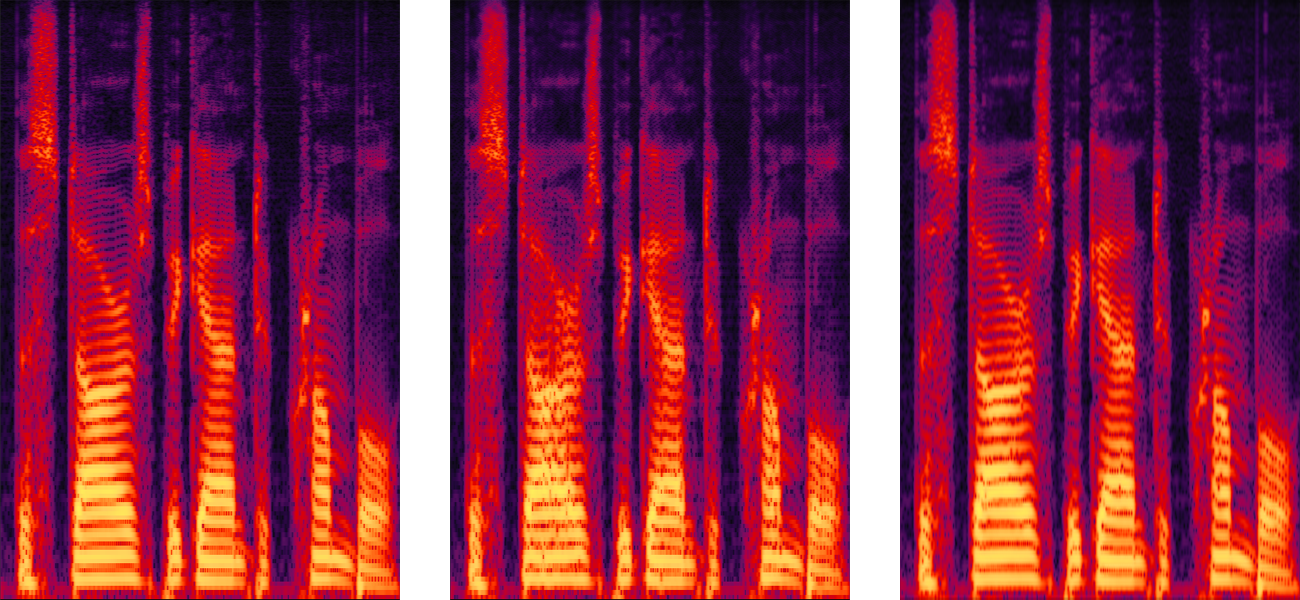}
    \caption{Spectrograms of a clean audio and audios generated by WaveFM-6 Steps using the original STFT loss and our proposed STFT loss, from left to right.}
    \label{STFT}
\end{figure}

Ablation studies demonstrate that our multi-resolution STFT loss significantly improves sample quality in one-step generation for WaveFM. This is further illustrated in \autoref{STFT}, where the spectrograms of audio generated using the original loss exhibit lower accuracy compared to those generated using our proposed loss.

The second auxiliary loss function is the average L1 loss of the mel-spectrograms. Both subjective and objective experiment results indicate that these two auxiliary losses significantly improve sample quality. Besides, we provide the Pytorch implementation details of our multi-resolution STFT loss in \autoref{sec:appd}.

It is worth noting that our reparameterization diverges at time $t=1$. Therefore, during distillation, we restrict the range of $t$ to $[0, 0.99]$, as a $t$ that is too large is ineffective since the waveforms are already sufficiently clean. The training process is summarized in \textbf{Algorithm} \ref{algo:train}, where $\operatorname{prior}(\bm{m})$ denotes the diagonal covariance matrix derived from the mel-spectrogram $\bm{m}$. For simplicity, we denote the reparameterized $\bm{v}^\prime$ as $\bm{v}_{\bm{\theta}}$ in the algorithm.

\begin{algorithm}[h]
   \caption{Train WaveFM}
   \label{algo:train}
\begin{algorithmic}
    \STATE {\bfseries Input:} 
neural network $\bm{v}_{\bm{\theta}}$, mel-spectrogram $\bm{m}$, time step $t \sim U[0,1]$, $\lambda_0=0.02,\lambda_1=0.02$
    \REPEAT
    \STATE $\bm{x}_1\sim p_{\operatorname{data}}(\bm{x}|\bm{m}),\bm{x}_0 \sim \mathcal{N}(\mathbf{0},\operatorname{prior}(\bm{m}))$
    \STATE $\bm{x}_t=t\bm{x}_1+(1-t)\bm{x}_0,\bm{v}_1=\bm{v}_{\bm{\theta}}(\bm{x}_t,t,\bm{m})$
    \STATE $\operatorname{Loss}=\frac{1}{\operatorname{min}(0.1,1-t)}\left\lVert\bm{x}_1-\bm{v}_1\right\rVert^2$
    \STATE $\qquad+\lambda_0\operatorname{STFTLoss}(\bm{x}_1,\bm{v}_1)$
    \STATE $\qquad+\frac{\lambda_1}{M} \left\lVert\operatorname{mel}(\bm{x}_1)-\operatorname{mel}(\bm{v}_1)\right\rVert_1$
    \STATE  Update $\bm{\theta}$ via gradient descent 
    \UNTIL WaveFM converges
\end{algorithmic}
\end{algorithm}

\subsection{Distillation Objective}

The conventional inference method using numerical ODE solvers typically requires numerous steps to generate waveforms, which contradicts our efficiency demand. Inspired by consistency distillation \citep{song2023consistency} for SDEs, we propose a specialized consistency distillation algorithm for our model, summarized in \textbf{Algorithm} \ref{algo:distill}.

\begin{algorithm}[h]
   \caption{Distill WaveFM}
   \label{algo:distill}
\begin{algorithmic}
    \STATE {\bfseries Input:} 
     student network $\bm{v}_{\bm{\theta}}$, teacher network $\bm{v}^\prime_{\bm{\theta}^\prime}$, EMA decay rate $\mu=0.999$,
     mel-spectrogram $\bm{m}$, distance $d(\cdot,\cdot)$, time duration $\Delta t=0.01$
    \STATE \textbf{Initialize} EMA parameters $\bm{\theta}^-=\bm{\theta}$
    \REPEAT
    \STATE $\bm{x}_1\sim p_{\operatorname{data}}(\bm{x}|\bm{m}),\bm{x}_0 \sim \mathcal{N}(\mathbf{0},\operatorname{prior}(\bm{m}))$
    \STATE $t\sim\widetilde{\mathcal{N}}\left(0,0.33^2\right)$, where $\widetilde{\mathcal{N}}$ refers to the truncated Gaussian distribution into $[0,0.99]$
    \STATE $\bm{x}_t=t\bm{x}_1+(1-t)\bm{x}_0$
    \IF{$t+\Delta t>0.99$}
        \STATE $\operatorname{target}=\bm{x}_1$
    \ELSE
        \STATE$\bm{x}_{t+\Delta t}=\operatorname{ODESOLVE}(\bm{v}^\prime_{\bm{\theta}^\prime},\bm{x}_t,t,\Delta t)$
        \STATE $\operatorname{target}=\bm{v}_{\bm{\theta}^-}(\bm{x}_{t+\Delta t},t+\Delta t,\bm{m})$
    \ENDIF
    \STATE Loss $=d(\bm{v}_{\bm{\theta}}(\bm{x}_t,t,\bm{m}),\text{target})$
    \STATE Update $\bm{\theta}$ via gradient descent 
    \STATE $\bm{\theta}^-=\operatorname{stopgrad}(\mu \bm{\theta}^- +(1-\mu)\bm{\theta}) $
    \UNTIL WaveFM converges
\end{algorithmic}
\end{algorithm}

In the algorithm, the student network is initialized with the pretrained model, and we calculate the exponential moving average (EMA) of the student network parameters to produce the consistency training target, which is essential for stabilizing the distillation procedure. During the distillation process, we sample $t$ from a truncated $\mathcal{N}\left(0,0.33^2\right)$ within the range $[0,0.99]$, rather than from $\mathcal{U}(0,0.99)$. This choice is made because the error at time steps near $t=0$ is more critical in one-step generation. The distance function $d(\cdot,\cdot)$ in our algorithm serves as shorthand for the training loss between network outputs and targets, comprising four terms as previously mentioned. The $\operatorname{ODESOLVE}$ can utilize any numerical solver, and we employ the Euler method in our implementation. Notably, since we reparameterize the original objective, it is necessary to reconstruct the original function for the numerical ODE solver. Furthermore, instead of parameterizing the consistency function in a continuously differentiable manner as in \citep{song2023consistency}, we directly set the targets to be clean audios at time points close to 1 and use the neural network to predict results at smaller time points. This represents a significant deviation from traditional consistency models, as the quality of generated audio would severely degrade if we parameterized our model conventionally. Additionally, our method is compatible with our dynamic prior distribution, requiring only sampling from it, in contrast to the original consistency models that rely on a fixed prior distribution determined by the solutions to their forward SDEs.

\subsection{Network Architecture}
\label{sec:appb}

\begin{figure}[ht]
\centering
\includegraphics[width=\linewidth]{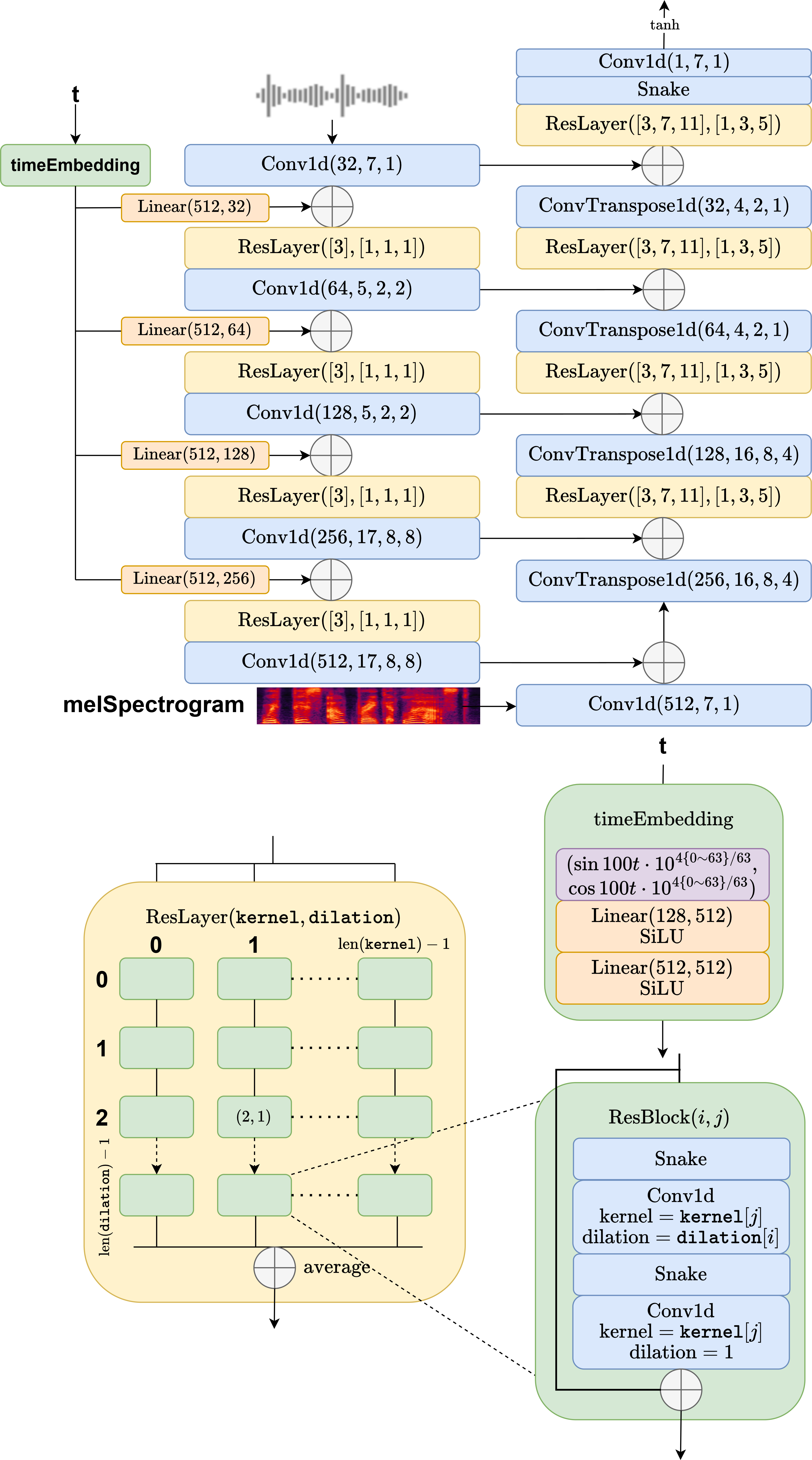}
\caption{Network architecture. \texttt{Conv1d} and \texttt{ConvTranspose1d} are set with parameters (output channel, kernel width, dilation, padding).}
\label{model}
\end{figure}

The function \(\bm{v}(\bm{x},t)\) is predicated on an asymmetric U-Net model of 19.5M parameters, adopting multi-receptive field modules, which are first introduced in the Hifi-GAN \citep{kong2020hifi} generator, referred to as ResBlocks and ResLayers in \autoref{model}. 
In ResBlock \texttt{Conv1d} takes ``same'' padding. Each ResLayer is defined with a kernel list and a dilation list, and their outer product defines both the ResBlock matrix, and the kernel width and dilation of convolutional layers in each ResBlock. 

Given that mel-spectrograms provide detailed conditions, the neural network's primary task is to upsample the mel-spectrogram and refine the waveform step by step. Consequently, we allocate most of our parameters and FLOPs to the upsampling process of the U-Net rather than distributing them equally between both sides. Additionally, we employ dilated convolutional layers with larger kernel sizes in the upsampling process to reconstruct audio from mel-spectrograms, while the downsampling process features simpler convolutional layers. On the left column of \autoref{model} are downsampling ResLayers, each containing a $4\times 1$ ResBlock matrix, while on the right columns are upsampling ResLayers, each containing a $3\times 3$ ResBlock matrix, following the structure from Hifi-GAN. In each ResBlock the number of channels remains unchanged from layers to layers. 

Inside the multi-receptive field modules, we adopt the snake-beta activation function from BigVGAN, defined with channel-wise log-scale parameters $\alpha$ and $\beta$:
\begin{equation}
    \operatorname{snake}(x)=x+\frac{1}{e^\beta+\epsilon} \sin^2(e^\alpha x),
\end{equation}
where $\epsilon=10^{-8}$ to ensure numerical stability. Note that this activation function, with its periodic inductive bias, is ineffective for the original flow matching model, as the noise in the random derivatives disrupts periodic patterns in the audio.

The downsampling and upsampling processes are implemented using strided and transposed convolutions, respectively. This design choice reflects our goal of generating waveforms directly from mel-spectrograms, where the additional information from downsampling features is less critical and primarily serves as a controller for the upsampling process.

For time representation, we follow \citep{kong2020diffwave} by embedding $t \in [0,1]$, scaled by 100 to align its magnitude with diffusion models, into a 128-dimensional positional encoding vector
\begin{equation}
\begin{aligned}
    \Big[&\sin\left(100t\cdot10^{\frac{0\times4}{63}}\right),\dots,\sin\left(100t\cdot10^{\frac{63\times4}{63}}\right),\\
        &\cos\left(100t\cdot10^{\frac{0\times4}{63}}\right),\dots,\cos\left(100t\cdot10^{\frac{63\times4}{63}}\right)\Big]
\end{aligned}
\end{equation}
These 128-dim time embeddings are first expanded to 512-dim after two linear-SiLU layers, then reshaped to the desired shape of each resolution, and finally added to the hidden layers during the downsampling process.

\section{Experiments}
\begin{table*}[ht]
\centering
\begin{tabular}{c|cccccc}
\hline
    \textbf{Model}&\textbf{SMOS ($\uparrow$)}&\textbf{M-STFT ($\downarrow$)}&\textbf{PESQ ($\uparrow$)}&\textbf{MCD ($\downarrow$)}&\textbf{Period ($\downarrow$)}&\textbf{V/UV F1 ($\uparrow$)}\\
\hline
\text{Ground Truth} &4.41$\pm$0.06
& 0.000 & 4.644 & 0.000
& 0.000 & 1.000 \\ 
\text{Hifi-GAN V1} &4.09$\pm$0.08
& 0.995& 2.943& 1.942& 0.163& 0.928\\ 
\text{Diffwave-6 Steps} &4.07$\pm$0.09
& 1.279& 2.956& 2.675& 0.154 & 0.936 \\ 
\text{PriorGrad-6 Steps} &4.12$\pm$0.10& 1.832& 3.161& 2.519& 0.159 & 0.937 \\ 
\text{FreGrad-6 Steps} &4.08$\pm$0.09
& 1.893& 3.148& 2.573& 0.165 & 0.932 \\ 
\text{FastDiff-6 Steps} &4.06$\pm$0.08& 2.181& 2.889& 3.264& 0.156 & 0.937 \\ 
\text{BigVGAN-base} &4.17$\pm$0.09
& 0.876 & 3.503& 1.316& 0.130& 0.945 \\ 
\hline
\text{WaveFM-1 Step} &4.11$\pm$0.08
& 0.872& 3.514& 1.355& 0.141& 0.943\\ 
\text{WaveFM-6 Steps} &\textbf{4.19}$\pm$0.10& \textbf{0.841}& \textbf{3.882}& \textbf{1.150}& \textbf{0.116}& \textbf{0.956}\\ 
\hline
\end{tabular}

\caption{Subjective results with 95\% confidence interval and objective evaluation results on LibriTTS dev set.}
\label{tab:libritest}
\end{table*}

\begin{table*}[ht]
\centering
\begin{tabular}{c|cccccc}
\hline
\textbf{Model} &\textbf{SMOS ($\uparrow$)}& \textbf{M-STFT ($\downarrow$)} & \textbf{PESQ ($\uparrow$)} & \textbf{MCD ($\downarrow$)} & \textbf{Period ($\downarrow$)} & \textbf{V/UV F1 ($\uparrow$)} \\ 
\hline
\text{Ground Truth} &4.38$\pm$0.08
& 0.000 & 4.644 & 0.000 & 0.000 & 1.000 \\ 
\text{Hifi-GAN V1} &3.81$\pm$0.11
& 1.288& 2.635& 2.469& 0.182& 0.924\\ 
\text{Diffwave-6 Steps} &3.85$\pm$0.10
& 1.354& 2.731& 3.875& 0.171& 0.929\\ 
\text{PriorGrad-6 Steps} &3.92$\pm$0.09& 1.925& 3.156& 3.439& 0.168& 0.931\\ 
\text{FreGrad-6 Steps} &3.87$\pm$0.10
& 1.960& 2.953& 3.325& 0.180& 0.924\\ 
\text{FastDiff-6 Steps} &3.79$\pm$0.08
& 2.257& 2.659& 4.331& 0.179& 0.923\\ 
\text{BigVGAN-base} &3.95$\pm$0.09
& 1.262& 3.170& 1.597& 0.151& 0.942\\ 
\hline
\text{WaveFM-1 Step} &3.96$\pm$0.09& 1.034& 3.287& 1.589& 0.148& 0.947\\ 
\text{WaveFM-6 Steps} &\textbf{4.05}$\pm$0.08& \textbf{0.968}& \textbf{3.639}& \textbf{1.342}& \textbf{0.135}& \textbf{0.952}\\ 
\hline
\end{tabular}
\caption{Subjective results with 95\% confidence interval and objective evaluation results on MUSDB18-HQ.}
\label{tab:mixture}
\end{table*}

\subsection{Datasets}

To ensure fair and reproducible comparisons with other competing methods, we employ the LibriTTS dataset \citep{zen2019libritts}, a large-scale corpus of read English speech comprising over 350,000 audio clips at 24,000 Hz, spanning approximately 1,000 hours of recordings from multiple speakers. All models are trained using the full dataset, including train-clean-100, train-clean-360, and train-other-500. For our mel-spectrograms, we generate 100-band mel-spectrograms with a frequency range of [0, 12] kHz. The FFT size, Hann window size, and hop size are set to 1024, 1024, and 256, respectively.

To evaluate the model’s ability to generalize in out-of-distribution scenarios, we use the MUSDB18-HQ music dataset \citep{rafii2017musdb18}. This multi-track dataset includes original mixture audio, along with four separated tracks: vocals, drums, bass, and other instruments.

\subsection{Training and Evaluation Metrics}

The detailed architectures and configurations of the models can be found in \autoref{sec:appb}. For training, the model is run on a single NVIDIA RTX 4090 GPU, starting with an initial learning rate of $7.5\times 10^{-5}$ and a batch size of 16. The learning rate decays according to a cosine annealing schedule, with the final learning rate set to $5\times 10^{-6}$, and the training process spans 1 million steps. 
We adopt the AdamW optimizer, setting the betas to (0.9, 0.99) and the weight decay rate to $5\times 10^{-4}$. The distillation stage, however, only requires 25,000 steps, with the initial learning rate reduced to $2\times 10^{-5}$. During distillation, we adjust the weight decay rate to $1\times 10^{-2}$ and set the betas to (0.8, 0.95). The time duration during distillation is set to 0.01. The training process requires approximately two days, while the distillation process is completed in about two hours. 
Given the need to evaluate performance on out-of-distribution data, we conduct a 5-point Similarity Mean Opinion Score (SMOS) test as described in BigVGAN \citep{lee2022bigvgan}. This subjective evaluation is carried out by ten volunteers, and the reported SMOS scores include a 95\% confidence interval. To ensure evaluation accuracy, 150 audio samples are generated per dataset for testing, with six different workers rating each sample.

Additionally, we incorporate objective automatic metrics to assess sample quality, including Multi-resolution STFT (M-STFT) loss \citep{yamamoto2019parallel}, Perceptual Evaluation of Speech Quality (PESQ) \citep{rix2001perceptual}, Mel-cepstral distortion (MCD) with dynamic time warping \citep{kubichek1993mel}, Periodicity error, and the F1 score for voiced/unvoiced (V/UV) classification \citep{morrison2021chunked}. Details on the implementation of these metrics are provided in \autoref{sec:appc}
. 

Moreover, we compute the real-time factor (RTF) using the same RTX 4090 GPU, defined as the ratio of total generated audio duration to inference time. It is important to note that the inference time excludes data loading and saving times.

\begin{figure*}[t]

    \centering
    \includegraphics[width=1.0\linewidth]{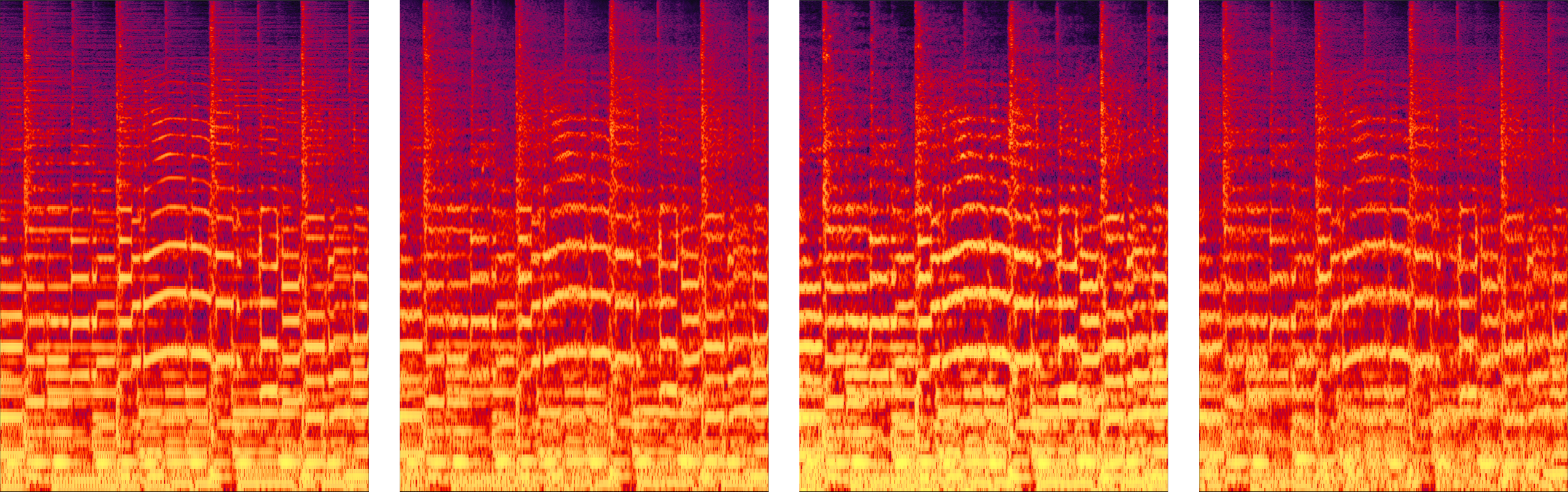}
    \caption{Spectrograms of a music clip (Ground Truth, WaveFM-6 Steps, PriorGrad-6 Steps, BigVGAN-base)}
   \label{music}
\end{figure*}
\subsection{Comparison With Other Models}
We conduct a series of experiments on speech synthesis tasks to evaluate our model. Models we have compared with are listed below: 

\textbf{Hifi-GAN V1} \citep{kong2020hifi}, \textbf{BigVGAN-base} \citep{lee2022bigvgan}, two well-known GAN vocoders; \textbf{DiffWave} \citep{kong2020diffwave}, \textbf{PriorGrad} \citep{lee2021priorgrad}, \textbf{FastDiff} \citep{huang2022fastdiff}, \textbf{FreGrad} \citep{nguyen2024fregrad}, four diffusion probabilistic models: all proved to be high-fidelity. We train these models for 1M steps with a batch size of 16 on LibriTTS following the setups as in the original papers.

The results in \autoref{tab:libritest} show that our models demonstrate superior performance over various previous models in terms of sample quality. In the table, WaveFM-1 Step and WaveFM-6 Steps audios are generated by our distilled model and undistilled with 6 steps Euler solver respectively. According to the data, our 6 steps model achieves the best performance on M-STFT, PESQ, MCD, Periodicity error and V/UV F1 score and our consistency distillation algorithm doesn't compromise too much sample quality in order to achieve one step generation. 

Besides, our model has advantages in terms of synthesis speed. The RTF results in \autoref{tab:rtf} have shown that our distilled model enjoys a inference speed close to Hifi-GAN V1, which is much higher than the previous diffusion models since they need several steps to generate waveforms.
\begin{table}[h]
\centering
\begin{tabular}{c|c}
\hline
    \textbf{Model}&\textbf{RTF ($\uparrow$)}\\
\hline
    Hifi-GAN V1&\textbf{325}$\times$\\
    Diffwave-6 Steps&16.7$\times$\\
    PriorGrad-6 Steps&16.7$\times$\\
    FreGrad-6 Steps&35.3$\times$\\
    FastDiff-6 Steps&69.4$\times$\\
    BigVGAN-base&90.3$\times$\\
\hline
    WaveFM-1 Step& \textbf{303}$\times$\\
    WaveFM-6 Steps& 50.2$\times$\\
\hline
\end{tabular}
\caption{Real Time Factor (RTF)}
\label{tab:rtf}
\end{table}
\subsection{Out-of-Distribution Situation}
We demonstrate the generalizability of WaveFM 
using the musical dataset MUSDB18-HQ.
The SMOS test is conducted by uniformly sampling from the five tracks: drums, bass, vocal, others, and mixture. Since automatic evaluators are primarily designed for speech analysis, we use vocal track samples for automatic evaluation, where audio segments with high silence ratios are removed.
The results in \autoref{tab:mixture} indicate our model exhibits 
commendable performance in unseen scenarios,
exceeding the performance of the baseline models. 
To be specific, WaveFM-6 Steps model performs significantly better than previous diffusion models, and even distilled WaveFM-1 Step model can generate acceptable waveforms compared to other methods,
which implies that our models generalize well on out of distribution data.
We can further illustrate this point by visualizing the spectrograms of the generated audio. \autoref{music} shows that our model's spectrogram is closer to ground truth spectrogram, and comparing to our model, PriorGrad tends to overestimate low frequency components and underestimate high frequency components, while BigVGAN-base fails to generate the regular 
components in the clean spectrogram neatly. Figuratively speaking, 
PriorGrad's spectrogram is too light at the bottom and a little dark at the top, and BigVGAN's spectrogram fails to keep the regular lines that can be found in clean spectrogram.

\subsection{Ablation Study}

\begin{table}[h]
    \centering
    \begin{tabular}{c|c}
    \hline
    \textbf{Model}       & \textbf{SMOS ($\uparrow$)}  \\
    \hline
    Ground Truth         & $4.41 \pm 0.06$ \\
    WaveFM-1 Step            & $\mathbf{4.11 \pm 0.08}$ \\
    w/o Snake Activation & $4.03 \pm 0.07$ \\
    w/o Conditioned Prior & $3.79 \pm 0.08$ \\
    with Original STFTLoss& $4.05\pm 0.07$\\
    w/o Auxiliary Losses & $3.97 \pm 0.09$ \\
    w/o Reparameterization & $3.88 \pm 0.07$ \\
    \hline
    \end{tabular}    
    \caption{Ablation study results on LibriTTS Test set.}
    \label{tab:abulation}
\end{table}

In order to show that our structural designs are effective, we have conducted several ablation studies on LibriTTS Test set. Here are the observations:
\begin{itemize}
\item[1.]After removing the random derivative term, the snake activation function with periodic inductive bias, as used in BigVGAN, improves the sample quality of WaveFM;
\item[2.]The mel-conditioned prior distribution significantly improves the one-step sample quality, indicating that reducing unnecessary distribution transportation costs is effective;
\item[3.]For our model, replacing the spectral convergence loss $L_{\text{sc}}$ by the phase angle loss $L_{\text{pha}}$ and incorporating gradient and Laplacian operators can improve
sample quality, which is consistent with the visual results in \autoref{STFT}.
\item[4.]The auxiliary losses are important to improve the sample quality, which aligns with the experiment results of GANs. Moreover, the results suggest that additional losses cannot be efficiently applied to the original flow matching objective unless it is reparameterized to directly predict waveforms; thus, the reparameterization is indeed crucial. 
\end{itemize}

\section{Conclusion}
We propose WaveFM, a high-fidelity vocoder for speech synthesis conditioned on mel-spectrograms. First, it utilizes the energy information from mel-spectrograms to generate a prior distribution with low variance. Additionally, reparameterizing the original flow matching objective not only introduces periodic inductive bias into the neural network, but also enables the inclusion of auxiliary losses. Specifically, we design a multi-resolution STFT loss function to enhance sample quality for our model. Finally, our consistency distillation algorithm allows WaveFM to produce audio in one step without significantly sacrificing sample quality. Together, these techniques greatly improve both the quality and efficiency of WaveFM, with SMOS tests and automatic evaluators confirming that it performs competitively against previous diffusion models.

\section{Limitations and Potential Risks}

The main limitation of our work lies in the noticeable gap in sample quality between multi-step and single-step models. While single-step models offer faster synthesis, their quality still lags behind, which restricts their usability for generating high-quality waveforms at scale. This trade-off remains a challenge, as improving the performance of single-step models without resorting to adversarial training requires further exploration and insights.

While our proposed model improves the accessibility of high-fidelity speech synthesis, it also introduces potential risks. By lowering the technical barriers, our approach could inadvertently facilitate misuse, such as more convincing voice spoofing or impersonation in media, customer service, or telephone scams. This raises concerns about the ethical implications of deploying such technology without proper safeguards. 
\bibliography{custom}
\newpage
\appendix

\section{Proof of Theorem 1}
\label{sec:appa}
\paragraph{Theorem 1} Let $\bm{x}_t$ be a continuously differentiable random
process on $t\in[0,1]$ and $p(\bm{x},t)$ be its probability density function. 
We denote the prior distribution as $\bm{x}_0$ and the ground truth distribution as $\bm{x}_1$,
if the conditional expectation 
$\mathbb{E}\left[\frac{\operatorname{d}\!\bm{x}_t}{\operatorname{d}\! t}\Big|\bm{x}_t=\bm{x}\right]$ is locally Lipschitz, let 
\begin{equation}
  \bm{v}(\bm{x}_t,t)=\mathbb{E}\left[\frac{\operatorname{d}\!\bm{x}_t}{\operatorname{d}\!t}\bigg|
  \bm{x}_t=\bm{x}\right],
\end{equation}
then we can draw samples from data distribution $\bm{x}_1$
by drawing samples from prior distribution  $\bm{x}_0$ and then 
solving the following ODE with an initial value $\bm{x}_0$ at time $t=0$.
\begin{equation}
    \frac{\operatorname{d}\!\bm{x}}{\operatorname{d}\!t}=\bm{v}(\bm{x},t)
\end{equation}
\paragraph{Proof} To generate data by drawing samples from prior distribution  $\bm{x}_0$ and then 
solving the following ODE with an initial value $\bm{x}_0$ at $t=0$:
\begin{equation}
    \frac{\operatorname{d}\!\bm{x}}{\operatorname{d}\!t}=\bm{v}(\bm{x},t),
\end{equation}
we only need to check that the probability density function $p(\bm{x},t)$
satisfies the probability flow equation below with given $\bm{v}(\bm{x},t)$
since when solving an ODE with locally Lipschitz condition from time $t=0$ to $t=1$ with initial value $\bm{x}_0$, the evolution of 
probability density is described by the probability flow equation.
\begin{equation}
   \frac{\partial p}{\partial t}
   (\bm{x},t)+\nabla_{\bm{x}}(p(\bm{x},t)\bm{v}(\bm{x},t))=0
\end{equation}
We can prove this by multiplying any finite supported continuously differentiable function $h(\bm{x})$ and then apply integral:
\begin{equation}
   \int\!h(\bm{x})\!\left(\!\frac{\partial p}{\partial t}
   (\bm{x},t)+\nabla_{\bm{x}}(p(\bm{x},t)\bm{v}(\bm{x},t))\!\right)\!\operatorname{d}\!\bm{x}=0
\end{equation}
\begin{equation}
\begin{aligned}
    &\frac{\partial}{\partial t}\int h(\bm{x})p(\bm{x},t)\operatorname{d}\!\bm{x}\\=&
   -\int h(\bm{x})\nabla_{\bm{x}}(p(\bm{x},t)\bm{v}(\bm{x},t)))\operatorname{d}\!\bm{x}
\end{aligned}
\end{equation}
Integrating by parts to the right hand side, since $h$ is finite supported, we have

\begin{equation}
\begin{aligned}
   &\frac{\operatorname{d}}{\operatorname{d}\!t}\int h(\bm{x})p(\bm{x},t)\operatorname{d}\!\bm{x}\\=&
   \int (p(\bm{x},t)\bm{v}(\bm{x},t))^\top \nabla_{\bm{x}} h(\bm{x})\operatorname{d}\!\bm{x}
\end{aligned}
\end{equation}

\begin{equation}
   \frac{\operatorname{d}}{\operatorname{d}\!t}\mathbb{E}[h(\bm{x}_t)]=
   \mathbb{E}\left[\bm{v}(\bm{x}_t,t)^\top \nabla_{\bm{x}} h(\bm{x}_t)\right]
\end{equation}

\begin{equation}
   \mathbb{E}\left[\frac{\operatorname{d}\!\bm{x}_t}{\operatorname{d}\!t}^\top \nabla_{\bm{x}} h(\bm{x}_t)\right]=
   \mathbb{E}\left[\bm{v}(\bm{x}_t,t)^\top \nabla_{\bm{x}} h(\bm{x}_t)\right]
\end{equation}

By the tower property of expectation, we have

\begin{equation}
\begin{aligned}
    &\mathbb{E}\left[\mathbb{E}\left[\frac{\operatorname{d}\!\bm{x}_t}{\operatorname{d}\!t}\bigg\vert\bm{x}_t\right]^\top \nabla_{\bm{x}} h(\bm{x}_t)\right]\\=&
   \mathbb{E}\left[\bm{v}(\bm{x}_t,t)^\top \nabla_{\bm{x}} h(\bm{x}_t)\right]
\end{aligned}
\end{equation}
since we choose
\begin{equation}
   \bm{v}(\bm{x},t)=\mathbb{E}\left[\frac{\operatorname{d}\!\bm{x}_t}{\operatorname{d}\!t}\bigg\vert\bm{x}_t=\bm{x}\right].
\end{equation}
Thus, the equations hold for any finite supported continuous differentiable function $h$. We have
\begin{equation}
   \frac{\partial p}{\partial t}
   (\bm{x},t)+\nabla_{\bm{x}}(p(\bm{x},t)\bm{v}(\bm{x},t))=0,
\end{equation}
since we can arbitrarily choose $h$. Then \textbf{Theorem 1} is proved and
we can draw samples by solving the ODE numerically.

\section{Implementations of Metrics}
\label{sec:appc}

\textbf{M-STFT:} We use the implementation in \texttt{Auraloss} \citep{steinmetz2020auraloss} with codes from \url{https://github.com/csteinmetz1/auraloss}.\\
\textbf{PESQ:} We resample the audios from 24,000 Hz to 16,000 Hz and pick the wideband version of PESQ from \url{https://github.com/ludlows/PESQ}.\\
\textbf{MCD:} We use the implementation at \url{https://github.com/jasminsternkopf/mel_cepstral_distance} with DTW enabled.\\
\textbf{Periodicity and V/UV F1:} Both are provided in CARGAN \citep{morrison2021chunked} at \url{https://github.com/descriptinc/cargan}.
\newpage
\section{Pytorch Implementation of Our Multi-resolution STFT Loss}
\label{sec:appd}
\lstset{
    language=Python,
    backgroundcolor=\color{white},
    basicstyle=\ttfamily\footnotesize,
    showstringspaces=false,
    keywordstyle=\color{PineGreen},
    stringstyle=\color{purple},
    numbers=none,
    breaklines=false,
    breakatwhitespace=true,
    captionpos=b,
    escapeinside={\%*}{*)},
    emph={to,filterTime,filterFreq,filterLaplacian,getSTFTLoss,conv2d,reshape,mean,sin,log,squeeze,unsqueeze,view_as_real,stft,tensor,hann_window,pow,abs,len,sqrt,cos},        
    emphstyle=\color{olive},
    morekeywords={torch,nn,functional,F},
    literate=*{
        {0}{{\color{teal}0\color{black}}}{1}
        {1}{{\color{teal}1\color{black}}}{1}
        {2}{{\color{teal}2\color{black}}}{1}
        {3}{{\color{teal}3\color{black}}}{1}
        {4}{{\color{teal}4\color{black}}}{1}
        {5}{{\color{teal}5\color{black}}}{1}
        {6}{{\color{teal}6\color{black}}}{1}
        {7}{{\color{teal}7\color{black}}}{1}
        {8}{{\color{teal}8\color{black}}}{1}
        {9}{{\color{teal}9\color{black}}}{1}
        {.0}{{\color{teal}.0\color{black}}}{2}
        {conv2d}{{\color{olive}conv2d\color{black}}}{6}
        {atan2}{{\color{olive}atan2\color{black}}}{5}
        {.pad}{{.\color{olive}pad\color{black}}}{4}
        {def}{{\color{blue}def\color{black}}}{3}
        {for}{{\color{blue}for\color{black}}}{3}
        {import}{{\color{blue}import\color{black}}}{6}
        {input}{{\color{black}input}}{5}
    }
}
\begin{lstlisting}
import torch
import torch.nn.functional as F

def filterTime(input):
    input = F.pad(input,
        pad=(1,0,1,1), mode="constant")
    weight = torch.tensor([
        [-1.0, 1.0],
        [-2.0, 2.0],
        [-1.0, 1.0]]).to(
        input.device).reshape(1,1,3,2)/4
    deltaT = torch.conv2d(
        input.unsqueeze(1),
        weight=weight,
    )
    return deltaT.squeeze(1)

def filterFreq(input):
    input = F.pad(input,
        pad=(1,1,1,0),mode="constant")
    weight = torch.tensor([
        [-1.0, -2.0, -1.0],
        [1.0, 2.0, 1.0]]).to(
        input.device).reshape(1,1,2,3)/4
    deltaF = torch.conv2d(
    input.unsqueeze(1),weight=weight)
    return deltaF.squeeze(1)

def filterLaplacian(input):
    input = F.pad(input,
        pad=(1,1,1,1),mode="constant")
    weight = torch.tensor([
        [-1.0, -1.0, -1.0],
        [-1.0, 8.0, -1.0],
        [-1.0, -1.0, -1.0]]).to(
        input.device).reshape(1,1,3,3)/8
    laplacian = torch.conv2d(
    input.unsqueeze(1),weight=weight)
    return laplacian.squeeze(1)

def getSTFTLoss(answer, predict,
    fft_sizes=(1024, 2048, 512),
    hop_sizes=(128, 256, 64),
    win_lengths=(512, 1024, 256),
    window=torch.hann_window,
):
    loss = 0
    for i in range(len(fft_sizes)):
        ansStft = torch.view_as_real(
            torch.stft(
            answer.squeeze(1),
            n_fft=fft_sizes[i],
            hop_length=hop_sizes[i],
            win_length=win_lengths[i],
            window=window(
                win_lengths[i],
                device=answer.device),
            return_complex=True)
        )
        predStft = torch.view_as_real(
            torch.stft(
            predict.squeeze(1),
            n_fft=fft_sizes[i],
            hop_length=hop_sizes[i],
            win_length=win_lengths[i],
            window=window(
                win_lengths[i],
                device=predict.device),
            return_complex=True)
        )
        ansStftMag = (ansStft[...,0]**2
            + ansStft[...,1]**2)
        predStftMag=(predStft[...,0]**2
            + predStft[...,1]**2)

        magMin = 1e-6
        mask = (ansStftMag > magMin)&(
                predStftMag > magMin)

        ansStftMag = torch.sqrt(
                ansStftMag + magMin)
        predStftMag = torch.sqrt(
                predStftMag + magMin)

        ansStftPha = torch.atan2(
                ansStft[..., 1][mask],
                ansStft[..., 0][mask])
        predStftPha = torch.atan2(
                predStft[..., 1][mask],
                predStft[..., 0][mask])

        deltaPhase = (
            ansStftPha - predStftPha)
        loss += torch.atan2(
            torch.sin(deltaPhase),
            torch.cos(deltaPhase),
        ).abs().mean()
        loss += (ansStftMag.log()
            - predStftMag.log()
        ).abs().mean()

        ansStftMagDT = filterTime(
                ansStftMag)
        ansStftMagDF = filterFreq(
                ansStftMag)
        ansStftMagLap = filterLaplacian(
                ansStftMag)

        predStftMagDT = filterTime(
                predStftMag)
        predStftMagDF = filterFreq(
                predStftMag)
        predStftMagLap=filterLaplacian(
                predStftMag)

        loss += 4.0 * (ansStftMagDF
            - predStftMagDF
        ).pow(2).mean()
        loss += 4.0 * (ansStftMagDT
            - predStftMagDT
        ).pow(2).mean()
        loss += 2.0 * (ansStftMagLap
            - predStftMagLap
        ).pow(2).mean()

    return loss / len(fft_sizes)
\end{lstlisting}
\end{document}